\documentclass[english,a4paper,12pt]{article}
\usepackage[T1]{fontenc}
\usepackage[ansinew]{inputenc}
\usepackage{babel}
\usepackage[left=3cm,top=3cm,bottom=3cm,right=3cm]{geometry}
\usepackage{graphicx,amsmath,booktabs,color,times,mathptmx}
	\graphicspath{{../Figure/}}	
\usepackage{natbib}
	
	\bibpunct{(}{)}{;}{a}{}{,}	
\usepackage[font=small]{caption}	
\usepackage[all]{xy}
	
\let\oldtabular\tabular 
\renewcommand{\tabular}{\small\oldtabular}

\begin{document}
\title{Trust and Partner Selection in Social Networks: An Experimentally Grounded Model}
\date{}
\author{Riccardo Boero\thanks{Dipartimento di Scienze Economiche e Finanziarie ``G. Prato'', Universit\`{a} di Torino and GECS --- Research Group in Experimental and Computational Sociology},{ } Giangiacomo Bravo\thanks{Dipartimento di Scienze Sociali, Universit\`{a} di Torino, Collegio Carlo Alberto and GECS --- Research Group in Experimental and Computational Sociology. Corresponding author. Mail address: Dipartimento di Studi Sociali, Via S. Ottavio 50, 10124 Torino, Italy. Email: giangiacomo.bravo@unito.it}{ }  and Flaminio Squazzoni\thanks{Dipartimento di Studi Sociali, Universit\`{a} di Brescia and GECS --- Research Group in Experimental and Computational Sociology}}
\maketitle

\begin{abstract}
This paper presents an experimentally grounded model on the relevance of partner selection for the emergence of trust and cooperation among individuals. By combining experimental evidence and network simulation, our model  investigates the link of interaction outcome and social structure formation and shows that dynamic networks lead to positive outcomes when cooperators have the capability of creating more links and isolating free-riders. By emphasizing the self-reinforcing dynamics of interaction outcome and structure formation, our results cast the argument about the relevance of interaction continuity for cooperation in new light and provide insights to guide the design of new lab experiments.

\end{abstract}

\section{Introduction}

Recent experimental studies have explicitly acknowledged the relevance of trust for cooperation in human societies \citep[e.g.][]{Barrera2009,BDM1995,Boero2009c,Boero2009,Camerer2003,Keser2003}. From a sociological perspective, the problem with these studies is that they took into account stylized and highly unrealistic interaction structures, e.g., random coupled subjects, so that the structure of the game was nothing but a sequence of one-shot interactions. This was due to the standard laboratory approach of economic studies that is aimed to purse the potential effect of the embeddedness of agents in complex social structures. On the contrary, many formal models with a more concrete sociological content investigated exactly how structural embeddedness affected trust and cooperation, but  without referring to empirically verified assumptions on agents' behavior \citep[e.g.][]{Cohen2001,Pujol2005}.

This paper aims to study the link between interaction outcomes and social structure formation by combining experimental evidence and social network simulation through an agent-based model (from now ABM). Unlike influential studies that emphasized the relevance of interaction continuity for the emergence of cooperation \citep[e.g.][]{Axelrod2002,Cohen2001}, we investigate whether the robustness of cooperation depends on the capability of agents to select their partners and whether this might be influenced by the particular network  structure  where they are embedded in. For this purpose, the advantage of combining laboratory experiments and simulation models is twofold. On one hand, experiments provide sound and clean data on agents' behavior on which to build informative micro-macro models. Unfortunately, this is a not so common practice, since most cooperation models have been based on stylized theoretical assumptions rather than on specific empirical evidence so far. On the other hand, ABM can extend experimental evidence by providing for the impossibility of exploring complex interaction structures and long-term macro dynamics and evolution in the laboratory. In this mutual benefit, there is a cross-fertilization that could be largely  beneficial for sociological research \citep[e.g.][]{BS2005,Chmura2007,DalForno2004,Ebenhoh2008,Janssen2006d,Rauhut2009,Rouchier2003}.

In light of this inspiration, we have extended the scope of a standard experimental economics game, the repeated investment game, to situations that were impossible to explore in the laboratory by introducing the effects of complex network structures. Our main substantial result is that different static network structures do not lead to outcomes substantially different from random networks. A substantial increase in cooperation occurs only when partner selection and dynamic structure formation are introduced, which allow cooperators to break their links with free-riders and let them to benefit from an higher number of interactions.

Our results show that the robustness of cooperation among agents, rather than being favored by the stability of the interaction structure, as suggested by previous studies \citep[e.g.][]{Axelrod2002}, have more to do with the selective intelligence of agents when they give rise to networks based on people who like each other that isolate bad apples. Moreover, our experimentally-grounded results confirm the findings of \citet{Ehuiluz2005} model, where the cooperative equilibrium in a Prisoner's Dilemma played on a dynamic network was guaranteed by ``leaders'' (agents in central position) that played a crucial role in sustaining cooperation in the system. In fact, also in our case, the presence of a cooperative agent in a central position in dynamic networks had a crucial influence in shaping a structure favorable to cooperation composed of trustful and trustworthy agents.

The remaining of the paper is organized as follows. The next section presents the research background. Section \ref{sec:exp} describes the experiment used to gather the data. Section \ref{sec:est} presents the estimation of agents' parameters. Section \ref{sec:models} introduces the models and presents the results. Finally, Section \ref{sec:discussion} discusses our findings.

\section{Research background}\label{sec:background}

Most studies on cooperation assumed that individuals have no choice over which opponents they play \citep[e.g.][]{Axelrod1984,Axelrod1997}. This implied that, especially in large populations, cheaters may escape punishment by roving from partner to partner, thereby exploiting cooperators. The simulation literature has suggested that the stability and continuity of the interaction structure might  provide a solution to combat exploitation by cheaters in social interaction. Simulating an iterative prisoner's dilemma, \citet{Cohen2001} have suggested that even just the continuity of the interaction patterns can make cooperative regimes come true in hostile environments, since the ``shadow of the future'' established through it preserved the contexts in which cooperation strategies worked, allowing later variants and emulations to find a supportive environment. In light of this, \citet{Axelrod2002} have emphasized that the persistence of interaction patterns was enough to guarantee cooperation, thereby dispensing from more complex linkage pattern mechanisms.

Although important, these findings did not capture the full picture of cooperation in human societies and underestimated more powerful social mechanisms that can sustain it. A prominent social mechanism that permits cooperators to decrease the risk of being cheated in many real social interactions, by isolating cheaters, is partner selection. It is likely that in concrete social situations individuals have preferential choice and can opt out from given interactions \citep[e.g.][]{Slonim2008}. Partner selection is not just a mean for individuals to do good preferential choices, but it also provides an incentive to the counterparts to be reliable and committed to others so as to avoid social isolation \citep[e.g.][]{Ashlock1996}. For example, \citet{Joyce-etal2006} have built a simulation model largely inspired by Axelrod's tournament that demonstrated that a conditional association strategy, i.e., assuming that agents left partners who defected against them and stayed with cooperative partners, was better performing than the conditional altruism postulated by Axelrod with the Tit-for-Tat strategy. A similar finding on the relevance of contingent strategies capable of discriminating good and bad partners was also found in a simulation by \citet{Aktipis2004}.

These interactional aspects have been explained by the strategic uncertainty reduction and the ``commitment bias'' arguments, with the former emphasizing rational strategic aspects and the latter intrinsic other-regarding or emotional motivations. The first position has been stressed by \citet{Kollock1994} in an experimental study on market transactions, where the challenge to deal with information asymmetries and uncertainty about the others' trustworthiness made long-term interaction partners look more attractive than others. \citet{Podolny2001} suggested that investment bankers that operated in markets characterized by a higher degree of uncertainty were likely to interact with colleagues they have interacted with in the past. \citet{Gulati1995} found the same evidence in an empirical study on firm alliances. More recently, \citet{Beckman2004} have confirmed this in an empirical study on interlock and alliance networks for the 300 largest US firms during the 1988-1993 period, where they found that there was a strong correlation between the type of market uncertainty and the stable/unstable nature of partner selection networks: firms facing firm-specific uncertainty were less selective and more open to new ties than firms facing collective market uncertainty. \citet{Hauk2001} developed a simulation model of an iterated prisoner's dilemma where partner selection was a learning strategy for agents in uncertain situations, with agents that learn how to discriminate between trustworthy and untrustworthy partners and to reward/punish others as well. By revising strategies according to selected partners, cooperators learn how to avoid that cheaters' exploitative strategies throve in the population. In two experiments conducted in the US and Japan, \citet{Yagamashi1998}  have found that uncertainty promoted commitment formation between partners but, in particular, that individuals trusting more others tended to form committed relations less frequently than skeptics, since they relied more on generalized trust. 

The second position argued that the risk and uncertainty of exchange provided the opportunity for partners to demonstrate their trustworthiness so that behavioral commitments and trust signals can sustain cooperation in an informal and spontaneous way, without negotiations, bargaining and binding agreements as in most social exchange literature \citep[e.g.][]{Molm2000}. Recent simulation studies have suggested that partner selection can generate a ``commitment bias'', so that people' commitment to existing partners can increase beyond instrumental reasons, thanks to the strengths of intrinsic other-regarding motivations \citep{Back2006,Back2008}. \citet{Back2010} has confirmed this hypothesis creating a lab commitment-dilemma game based on a market exchange between subjects who played in artists selling paintings'roles and computer-simulated collectors buying them, in six locations of three countries, i.e., the Netherlands, China and the US. The findings have showed that the initial commitment of buyers had a positive effect on partner selection, even when controlling for uncertainty reduction and material benefits.

Although full of insights of great importance for sociological analyses of social interaction, these studies did not take into the due account the relevance of complex social structures, and in particular the consequences of partner selection within certain social structure configurations. Most of the experimental literature on partner selection have typically investigated the internal aspects of the selection decision, rather than the consequences of it over time for the emergence of trust and cooperation \citep[e.g.][]{Haruvy2006,Kagel2000}. Recently, \citet{Slonim2008} have investigated the relevance of partner selection for trust and altruism in the lab, by playing an investment game and a variant of the dictator game, where subjects'partners were exogenously imposed or endogenously chosen. Nevertheless, they underestimated the relevance of studying partner selection sociologically within network formation processes, that is by investigating how network structure and partner selection can influence each other. In fact, partner selection is to be viewed as a driving force for network formation, stability and change \citep{Beckman2004}.

At the same time, the abundant literature on network formation that explained important aspects of cooperation in social and economic situations was more analytically oriented than explicitly based on experimental evidence, did not take into account large social systems \citep[e.g.][]{Calvo2001,Dutta2005,Jackson2002,Jackson1996} and did not investigate the crucial role of partner selection as a driving force for the network dynamics \citep[e.g.][]{DeVos2001}. An interesting exception was \citet{Ehuiluz2005}, who elaborated a social network simulation based on a spatial prisoner's dilemma to investigate the co-determination of individual strategies and network, but without partner selection at the agent level and lab data for the modeling of agent decisions.

Our paper aims to fill this gap by combining experimental evidence and network simulation to study partner selection, trust and network configurations.  This is to understand relevant interaction aspects that is impossible to look at in the laboratory and to provide new insights to improve the sociological content of current economic laboratory experiments.

\section{The experiment}\label{sec:exp}

Our goal was to observe individuals' behavior in asymmetric trust situations, i.e., where a trustor has to choose whether placing trust or not in another individual (the trustee) who has a rational incentive to cheat \citep[177--180]{Coleman1990}. The  experimental economics standard for studying these situations was the investment game, a protocol based on \citet{BDM1995}. The game has been repeated many times using different samples and under a vast range of different conditions \citep[e.g][]{Cronk2007,Keser2003,King-Casas-etal2005,Knoch2009,Ortmann2000} and, recently, it has gained interest also among sociologists \citep[e.g.][]{Barrera2009,Boero2009,Buskens2008,Molm2000}.  

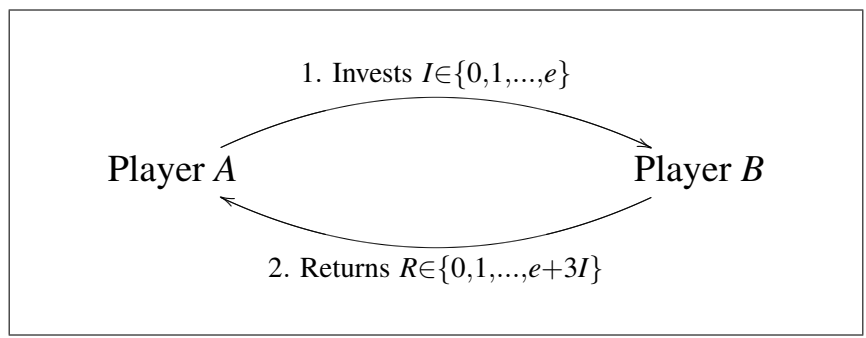
\begin{figure}[!t]%
\centering
\fbox{
\begin{minipage}{0.7\textwidth}
\large
\[
\xymatrix{
\text{Player}~A\ar@/^1cm/[rrrrr]^{1.~\text{Invests}~I \in \{0, 1,\ldots, e\}} & & & & & \text{Player}~B\ar@/^1cm/[lllll]^{2.~\text{Returns}~R \in \{0, 1,\ldots, e+3I\}}
}	
\]

\mbox{}
\end{minipage}
}
\caption{The investment game: player $A$ profit $= e - I + R$; player $B$ profit $= e + 3I - R$.}%
\label{fig:IG}%
\end{figure}

Figure  \ref{fig:IG} shows a schematic representation of a typical investment game.  Participants were coupled and randomly assigned to different roles, called player $A$ (the trustor) and player $B$  (the trustee). Both players received an endowment of $e$ ECU. First, player $A$ had to decide the amount $I$ between 0 and $e$ to send to $B$, keeping for him/herself the part $(e - I)$. 
The amount sent by $A$ was tripled by the experimenter and sent to the trustee, in addition to his/her own endowment, who  decided how much to return to $A$. As before, the amount returned by $B$ could be any integer between 0 and $(d + 3I)$. The amount $(d + 3I - R)$ not returned represented $B$'s profit, while $R$ was summed to the part kept by $A$ to form the final profit of the latter, which was hence calculated as $(e - I + R)$. All profits earned during the game were counted in experimental currency units (ECU), translated in real money using a fixed exchange rate and payed immediately after the end of the experiment.  The game is called ``investment game'' since the rule of  tripling the amount sent by the trustor implies that (a) the trustor deals with the uncertainty of paying a cost at the beginning of the interaction to gain possibly higher revenues at the end, and (b) the trustee has a return from the trustor's decision. 

Since $B$ had no rational incentive to return anything to $A$, the dominant strategy for $A$ players was to invest zero. This led to a unique subgame perfect equilibrium, where both players kept their entire endowments. However, this was inefficient since any sum grater than zero invested by $A$ was tripled by the experimenters and could lead to a Pareto superior outcome, with the optimum represented by $A$ investing the whole endowment. Despite these equilibrium predictions, a quite robust result of investment games played both in the laboratory and in the field was that $A$ players invested a substantial amount of their endowments (usually around 40\%)  and that $B$ players returned slightly less than the amount invested by $A$, although much less than a fair share of the overall amount $(d + 3I)$ that they received \citep[e.g][]{BDM1995,Boero2009,Cronk2007,Keser2003,Ortmann2000}.

Our experiment was based on a repeated version of the investment game as described above. One hundred and eight subjects were recruited through public announcements and played the game in six groups of 18 individuals. Half of the subjects were students of the University of Brescia, half of the University of Torino (Cuneo campus). The experiment took place in Spring 2009 in the computer laboratories of the Faculties of Economics of the two universities, which were both equipped with the experimental software z-Tree \citep{Fischbacher2007}. All groups received exactly the same instructions and played a repeated investment game using an identical computer interface. The endowment was 10 ECU with an exchange rate of 1 ECU = 2.5 Euro. Participants earned, on average, approximatively 15 Euros, including a show up fee of 5 Euros, that were paid immediately after the experiment. The experiment, including instruction reading, took approximatively one hour.

All interactions took place through the computer network and the subjects were unable to identifying their counterparts. Participants played a repeated investment game for a total of 10 periods and were informed in advance of the number of periods to play. The experiment used a ``stranger'' matching protocol, i.e., randomly re-coupling of subjects after each period. This protocol permitted to control for the use of direct reciprocity strategies among participants. The players' roles were randomly assigned at the beginning of the first period and subsequently changed on a regular basis, so that each subject played the same number of period in both roles.

The results show that subjects invested, on average, 3.48 ECU and returned 2.79 ECU, slightly less than what is usual in similar games.\footnote{All statistical analysis were conducted using the \emph{R} 2.10.1 platform \citep{R2009}.}  This can be explained if subject pool used, which was composed of undergraduate students in business and economics, is taken into account. This population is well known for under-average cooperation in experiments and this is even more true when subjects played investment and trust games \citep[see][65]{Camerer2003}.

\section{Parameter estimation} \label{sec:est}

We used these experimental data to calibrate an ABM that reproduced the behavior of the subjects. With regards to $A$ players, we estimated a coefficient $\beta_i$ that indicated how much the player modified his/her investment each round in function of the difference between the amount invested and the amount received by $B$ players in the previous round. For any player $i$ and round $t$, we computed the difference  $X_{it} = R_i - I_i$, where $I_i$ and $R_i$ were the amounts that $i$ invested and received as return from his/her investment in the previous round  respectively. We subsequently fitted the model
\begin{equation}
Y_{it} = \alpha_i + \beta_i X_{it} + \varepsilon
\label{eq:regression}
\end{equation}
where $Y_{it}$ was the amount invested by player $i$ in round $t$, in order to obtain two parameters $\alpha_i$ and $\beta_i$ for each subject that defined his/her behavior as $A$ player. The equation (\ref{eq:regression}) took into account that $A$ players could have had an individual constant propensity to trust represented by the individual intercept $\alpha_i$, but also the capability of reacting upon past experiences, which was captured in the  $\beta_i$ coefficient.
 
On the other hand, $B$ players were supposed to react mainly against what they received from $A$ players. In order to capture their behavior, we estimated a third coefficient $\gamma_i$ defined as the average amount returned by each subject as proportion of the amount received in each round plus the fixed endowment. Therefore, the parameter $\gamma_i$ represented an estimate of the player's trustworthiness.

\begin{figure}[!t]
\centering
\includegraphics[width=.33\textwidth]{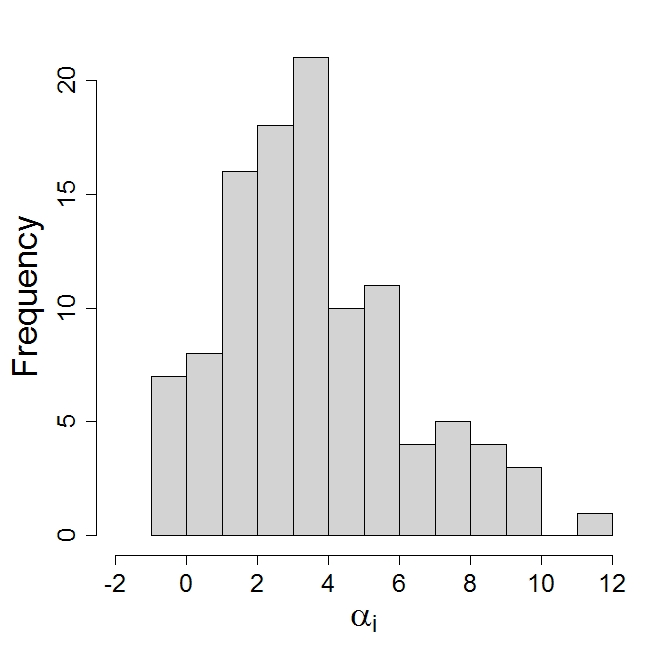}\includegraphics[width=.33\textwidth]{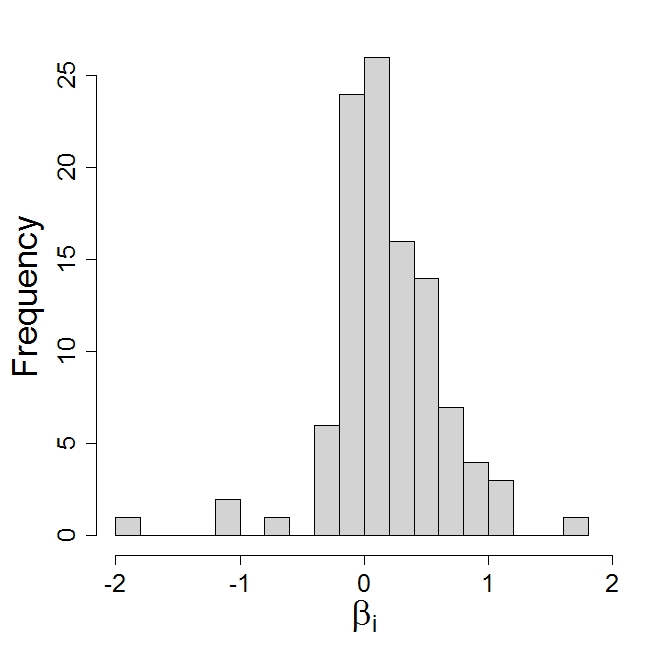}
\includegraphics[width=.33\textwidth]{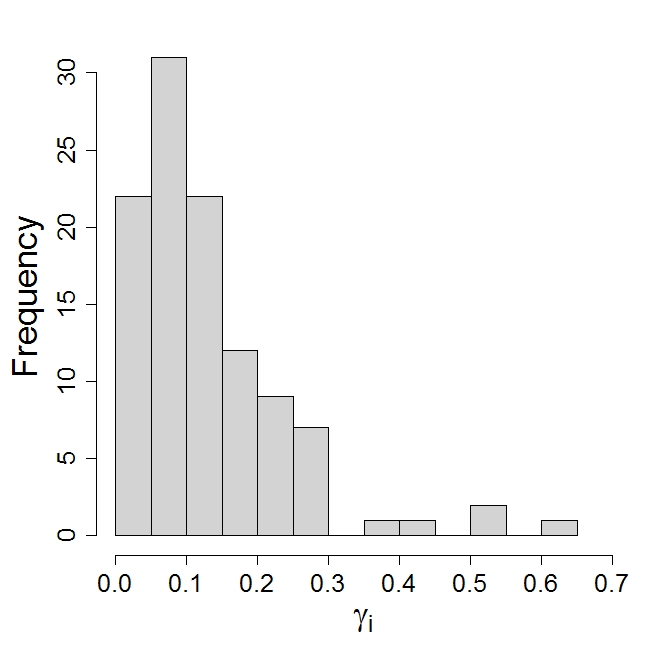}
\caption{Distribution of the estimated coefficients}
\label{fig:distr}
\end{figure}

We were able to successful estimate the parameter for 105 out of the 108 subject that participated in the experiment.  Figure \ref{fig:distr} presents the distribution of the estimated coefficients. It is worth noting that, while $\beta_i$ did not significantly correlated with the other parameters, the correlation between $\alpha_i$ and $\gamma_i$ was significant and positive ($r=0.44$, $p<0.001$). In other terms, trustful subjects tended to be trustworthy as well: a result consistent with the hypothesis that agents who have a general propension toward cooperation behave similarly in different situations.. 

\section{The models} \label{sec:models} 

\subsection{Experiment replication}

We first built an \emph{experimentLike} model that replicated the original experiment.\footnote{Note that we drawn from the data 105 valid estimations out of 108 subjects. Being the game played in dyads, we decided to remove randomly one of the experiment participants at the beginning of each run of the model, which, as a result, included only 104 agents.} In each period agents were coupled and played an investment game either as $A$ or as $B$ player. At the end of the periods they were randomly re-coupled while their roles alternated on a regular basis. The behavior of each agent depended on the coefficient estimated from the experimental data. More specifically, each agent $i$ playing in the role of $A$ in period $t$ invested an amount  
\begin{equation}
I(i,t) = \alpha_i + \beta_i \left[ R(i,t-1) - I(i,t-1) \right]
\label{eq:I}
\end{equation}
where $\alpha_i$ and $\beta_i$ were the coefficients of the corresponding participant while $I(i,t-1)$ and $R(i,t-1)$ were the amounts invested and received in the previous round. In the first round of the game, we set $I(i,0) = R(i,0) = 0$ that made agents choosing their first move on the basis of $\alpha_i$ alone. This was justified since $\alpha_i$ reflected the intrinsic propensity of the participants (and, henceforth, of our agents) to trust others players, while $\beta_i$ represented the effect of past experience. On the other hand, when playing on the $B$ positions agents returned 
\begin{equation}
R(i,t) = \gamma_i[3I(i,t) + 10]
\label{eq:R}
\end{equation}
which was simply their endowment multiplied by the individual parameter $\gamma_i$ estimated on the experimental data. 

Based on this setting, we built an \emph{experimentLike} model, which succeeded in reproducing the experimental data.\footnote{All models were programmed in Java using the JAS Library (http://jaslibrary.sourceforge.net/). Interested readers can obtain the source code upon request by the authors.} The average investment over 100 run of the \emph{experimentLike} model  was 3.57 ECU and the average return 2.76 ECU (see Tab. \ref{tab:means}). A $t$ test over individual investments/returns, averaged over the 10 periods of the game, confirmed that these figures did not significantly differ from the experiment (respectively, $t = 0.288$, $p = 0.774$ two sided, and $t = -0.085$, $p = 0.933$ two sided).

\subsection{The simulation scenarios}

Once checked the model capability of reproducing the experimental data, we designed new scenarios to test the plausible effects when important variables in the experimental design were modified. All scenarios used the experimental data as input and, more generally, were identical to \emph{experimentLike} model except for the  specific characteristic(s) under examination. Table \ref{tab:overview} provides an overview of all simulation scenarios.

\begin{table}[!t]
\centering
\begin{tabular}{lp{10cm}}
\toprule
Model name & Main characteristics \\
\midrule
\emph{experimentLike} &  $\bullet$ random coupling in each period\par $\bullet$ one way interaction \\
\midrule
\emph{twoWays} &  $\bullet$ random coupling in each period\par $\bullet$ two way interaction \\
\midrule
\emph{fixedCouples} &  $\bullet$ fixed couples\par $\bullet$ two way interaction \\
\midrule
\emph{denseNetwork} &  $\bullet$ fixed fully connected network\par $\bullet$ two way interaction \\
\midrule
\emph{smallWorld} &  $\bullet$ fixed small-world network\par $\bullet$ two way interaction \\
\midrule
\emph{scaleFree} &  $\bullet$ fixed scale-free network\par $\bullet$ two way interaction \\
\midrule
\emph{dynamic1Couples} & $\bullet$ dynamic network\par $\bullet$ broken links are replaced only for isolated agents\par $\bullet$ two way interaction\par $\bullet$ start from random coupling\\
\midrule
\emph{dynamic1Dense} &  $\bullet$ dynamic network\par $\bullet$ broken links are replaced only for isolated agents\par $\bullet$ two way interaction\par $\bullet$ start from dense network\\
\midrule
\emph{dynamic2Couples} &  $\bullet$ dynamic network\par $\bullet$ broken links are replaced only by one of the two formerly linked agents\par $\bullet$ two way interaction\par $\bullet$ start from random coupling\\
\midrule
\emph{dynamic2k10} &  $\bullet$ dynamic network\par $\bullet$ broken links are replaced only by one of the two formerly linked agents\par $\bullet$ two way interaction\par $\bullet$ start from from a regular network of degree 10\\
\bottomrule
\end{tabular}
\caption{The simulation scenarios}
\label{tab:overview}
\end{table}

A first simple variation of the game design was to increase the number of periods played by agents. This has been done for all scenarios, including the \emph{experimentLike} model, that were tested both using a 10 and a 30 period game. Then we explored the introduction of two way interaction among agents, i.e. in each period agents played both as $A$ and $B$ with their opponent  (\emph{twoWays} model). Note that, while this extension of the model could be of no particular interest in itself, it provided a baseline for subsequent scenarios which extended the interaction to a number of agents greater than two.

Since we knew that differences in the network structure can have crucial consequences for the cooperation at a macro level, we built four models that explored the effect of different network structures. The \emph{fixedCouples} model simply maintained the initial couples throughout the whole run. The \emph{denseNetwork} model introduced a fully connected network  where, in each period, every agent interacted with all other agents. In the \emph{smallWorld} model, we introduced a fixed ``small-world'' network structure \citep{Watts1998,Watts1999} with initial degree 4 and re-wiring probability 0.01. The \emph{scaleFree} model was based on a ``scale-free'' network  that conformed to \citet{Barabasi1999} algorithm with a degree equal to 3. In all these models, the network structure remained fixed throughout the whole game. Note that agents may have more than one interaction per period. More specifically, they played an investment game both as $A$ and as $B$ with all other agents with whom they were linked in the current period.

Finally, two models were introduced to investigate the effects of the introduction of a dynamic network in the game. Basically, these scenarios added a criterion of partner selection to the formation of couples, so that interaction structure was dynamically shaped by the outcome of the interaction \citep[e.g.][]{Corten2009,Ehuiluz2005,Flache2001}. Since there was no way to derive from the experiment the algorithm that agents used to break their links and create new ones, we identified certain plausible algorithms using as few assumptions as possible. We built two different models, both based on simple mechanisms but with a slightly different focus. The first one (named \emph{dynamic1}) was designed to avoid that any agent became isolated --- i.e. no longer linked to any other agent --- while the second (\emph{dynamic2}) was aimed at maintaining constant the total number of links in the network.

Both models were based on the idea that ``unsatisfied'' agents could break their links. In other words, we assumed that agents had a threshold happiness function so that they changed partners when they were unhappy. The threshold function was very simple: $A$ agents were happy when $B$ returns were higher or equal to the ones in the previous time step. This makes sense on the light of some fundamental attitudes of human beings as highlighted by economic psychology \citep[e.g.][]{KT2000}. Note also that a similar strategy for determining agents' satisfaction has been assumed, for instance, in \citet{Bravo2010} work that modeled the emergence of institutions in common-pool-resource situations. We assumed that a couple could be broken when at least one agent of the couple was unhappy. When there was no past information to evaluate the happiness --- i.e. in the first two periods --- the couple remained fixed. 

The only difference between the two dynamic models regarded the algorithm for the definition of the new links. In the \emph{dynamic1} model, broken links were replaced only when an agent became isolated. In this case a new link was formed between the isolated agent and a new agent different from the one with whom it was linked before. In the \emph{dynamic2} model, after each link break, one of the two formerly connected agents was chosen at random to initiate a new link. In this case, no check was made and agents may become isolated. This second algorithm maintained constant the number of links in the network, while changing its structure at the same time.

Both the \emph{dynamic1} and the \emph{dynamic2} model were tested starting from different initial network structures. More specifically, we used random coupling and a fully connected network as starting point for the former and random coupling and a random network of average degree 10 for the latter. The difference was due to the fact that the second  re-coupling algorithm determined a situation where each agent was connected to every other and there was no other possibility of re-linking than re-building the broken connection. Having two link replacement algorithms and two initial network structure, we  tested four dynamic models named, respectively, \emph{dynamic1Couples}, \emph{dynamic1Dense}, \emph{dynamic2Couples} and \emph{dynamic2k10}. All dynamic models were run for 30 periods in order to see the evolution of the network in a time longer than the original experiment.

\subsection{Static network model results}

In order to take into account the stochastic elements of the models, we ran 100 replications for each scenario. Table  \ref{tab:means} presents the average investments and returns in all models, comparing them with the experiment.

\begin{table}[!t]
	\centering
	\begin{tabular}{lcccc}
	\toprule
 & \multicolumn{2}{c}{10 period game} & \multicolumn{2}{c}{30 period game} \\
	\cmidrule(lr){2-3} \cmidrule(lr){4-5}
	Model name & $A$ investments & $B$ returns & $A$ investments & $B$ returns \\
	\midrule
 	\emph{experimentLike} & 3.57 (2.50) & 2.76 (2.62) & 3.56 (2.53) & 2.76 (2.63)\\
 	\emph{twoWays} &3.57 (2.52) & 2.76 (2.61) & 3.57 (2.54) & 2.76 (2.61) \\
 	\emph{fixedCouples} & 3.65 (2.53) & 2.91 (3.13)& 3.67 (2.56) & 2.92 (3.17) \\
 	\emph{denseNetwork} & 3.57 (2.54) & 2.76 (2.61) & 3.57 (2.54) & 2.76 (2.61)\\
 	\emph{smallWorld} & 3.58 (2.54) & 2.76 (2.62) &  3.57 (2.54) & 2.76 (2.62) \\
 	\emph{scaleFree} & 3.61 (2.54) & 2.80 (2.68) & 3.61 (2.54) & 2.80 (2.69) \\
	\midrule
	Experiment & 3.48 (2.69) & 2.79 (3.58) & -- & -- \\	
	\bottomrule			
	\end{tabular}
	\caption{Average investments and returns in the original experiment and in the simulation scenarios. Standard deviations are in parenthesis. Simulation results significantly different from the experimental ones are highlighted using a bold font.}
	\label{tab:means}
\end{table}

The first important point is that neither the modification of the network structure nor the increase of the number of periods changed significantly the outcome of the game. In most of the runs, models  reached their equilibrium before the 10\textsuperscript{th} period, maintaining it subsequently with no major change. Even the introduction of two-way interactions, which increased the number of investment games played by agents in every run, simply accelerated the approaching to the equilibrium without altering the results. Similarly, letting agents play over a fully connected network, where each agent interacted with all other agents in every single period (resulting in a total of 214,240 interactions per run in the 10 period game and in 642,720 interactions in the 30 period one) did not change the overall results. In all these cases, according to a $t$ test, the average investments and returns of the models did not  significantly differ from the experiment.

The introduction of networks where not all the agents had an equal number of interactions posed a challenge for our analysis. Up to now, to check the statistical significance of the difference between the simulation results and the experiment, we used a $t$ test over individual averages. This worked fine as long as all the agents interact an equal number of times, while it lead to an overestimation of the weight of agents with a lower number of interactions in case of networks where the number of links for each agent were different. For instance, the average investment in the 10 period \emph{scaleFree} model was 3.61 when considering each single interaction. However, taking in consideration the average contribution for each agent, the resulting average was 3.57. Although small, this difference could affect the significance of our results. Moreover, it was likely that the difference increased moving from static to dynamic networks. In order to take into account this issue, we performed our $t$ tests on the weighted means, where the weights were obviously given by the number of interactions performed by each agent.

The \emph{smallWorld} and \emph{scaleFree} model results were very similar to the previous ones: the average amount invested were slightly higher than the experiment, but a  $t$ test showed that this difference was not significant. Also the average returns were almost identical to the experiment.

An interesting question was whether the change in the network structure would have changed the relative advantage of cooperators (i.e. trustful and trustworthy individuals) over free-riders. We estimated the relation between the value of the three parameter defining each agent and the payoff earned in the different scenarios. Since  average investments and returns did not change moving from a 10 to a 30 period game,  we took into account only the 10 period game. Table \ref{tab:payoffs} reports the correlation coefficient between the agent parameters and its earnings.

\begin{table}[!t]
	\centering
	\begin{tabular}{lr@{}lr@{}lr@{}l}
	\toprule
	Model name & \multicolumn{2}{c}{$\alpha_i$} & \multicolumn{2}{c}{$\beta_i$} & \multicolumn{2}{c}{$\gamma_i$} \\
	\midrule
 	\emph{experimentLike} &  --0.75 & $^{***}$ & 0.08 & & --0.91 & $^{***}$\\
 	\emph{twoWays} & --0.74 & $^{***}$ & 0.10 & & --0.92 & $^{***}$ \\
 	\emph{fixedCouples} & --0.71 & $^{***}$ & 0.22 & $^{*}$ & --0.85 & $^{***}$ \\
 	\emph{denseNetwork} & --0.73 & $^{***}$ & 0.12 & & --0.92 & $^{***}$ \\
 	\emph{smallWorld} & --0.72 & $^{***}$ & 0.11 & & --0.91 & $^{***}$ \\
 	\emph{scaleFree} & --0.46 & $^{***}$ & 0.14 & & --0.69 & $^{***}$ \\
	\bottomrule			
	\end{tabular}
	\caption{Correlation between average payoffs per run and agents' parameters. Significance codes: $^{***} p < 0.001$, $^{**} p < 0.01$, $^{*} p < 0.05$.}
	\label{tab:payoffs}
\end{table}

In most models, earnings significantly and negatively correlated with both $\alpha_i$ and $\gamma_i$, while the correlation with $\beta_i$ was not significant. In other words, both trustful and trustworthy individuals earned lower payoffs in all games, while the agent ability to react to others' action did little difference. There was one noticeable exception to this rule, namely the \emph{fixedCouples} model where the correlation coefficient relative to $\beta_i$, although small, was positive and significant at the 5\% level. This means that reactive agents, by increasing their investments after a high return and decreasing them following a low return, were somewhat advantaged in this setting. This was probably due to the fact that the fixed couple structure of the game tended to favor reciprocity based strategies. Another partial departing from the rule mentioned above was the \emph{scaleFree} model. In this case the relation between the agents' trust (measured by $\alpha_i$) and their earnings was much weaker than in any other situation. To some extent, the same was true for their trustworthiness ($\gamma_i$). 

\subsection{Dynamic network model results}

Moving from a static to a dynamic network led to noticeable results (see Tab.	\ref{tab:meansDynamic}). Note that, being the network no longer static, the amount sent and returned in the different parts of the game varied. Therefore, we separately analyzed the results of periods 1--10, 11--20 and 21--30 of the game. The average cooperation in the \emph{dynamic1} models slightly increased, even if a $t$ test (performed on the weighted means for the reasons given above) showed that the difference was significant only for returns in the first ten rounds of the \emph{dynamic1Dense} model ( $t = 1.446$, $p = 0.075$, one sided). The \emph{dynamic1Dense} model resulted in higher cooperation levels than the \emph{dynamic1Couples} one because cooperative agents, starting from 103 links each,  were initially able to sever most of them, thereby avoiding be exploited by free-riders. However, this strategy became no longer possible once the situation settled down to an average of 1--2 links per agent, with the consequence that trustful agents' earnings decreased below those of the less trustful agents. 

\begin{table}[!t]
	\centering
	\begin{tabular}{lcccccc}
	\toprule
 & \multicolumn{2}{c}{Period 1--10} & \multicolumn{2}{c}{Period 11--20} & \multicolumn{2}{c}{Period 21--30} \\
	\cmidrule(lr){2-3} \cmidrule(lr){4-5} \cmidrule(lr){6-7}
	Model name & $A$ invest. & $B$ returns & $A$ invest. & $B$ returns & $A$ invest. & $B$ returns\\
	\midrule
 	\emph{dynamic1Couples} &  3.65 (2.58) &  2.92 (2.96) &  3.67 (2.60) & 2.95 (2.90) & 3.68 (2.62) & 2.96 (2.93) \\
 	\emph{dynamic1Dense} &  3.79 (2.67) &  \textbf{3.32} (3.20) & 3.66 (2.60) & 2.96 (2.96) & 3.68 (2.62) & 2.97 (2.94) \\
	 	\emph{dynamic2Couples} & 3.82 (2.68) & \textbf{3.37} (3.42) & \textbf{4.48} (3.01) & \textbf{5.02} (4.50) & \textbf{4.63} (3.11) & \textbf{5.58} (5.12) \\
	\emph{dynamic2k10} & \textbf{4.11} (2.82) & \textbf{4.00} (3.59) & \textbf{4.43} (3.01) & \textbf{4.85} (4.30) & \textbf{4.49} (3.04) & \textbf{5.02} (4.50) \\
	\midrule
	Experiment & 3.48 (2.69) & 2.79 (3.58) & -- & -- & -- & -- \\
	\bottomrule			
	\end{tabular}
	\caption{Average investments and returns in the original experiment and in the dynamic network models. Standard deviations are in parenthesis. Averages significantly different (at the 10\% level) from the experimental ones are marked in bold.}
	\label{tab:meansDynamic}
\end{table}

\begin{figure}[!t]
\centering
\includegraphics[width=.333\textwidth]{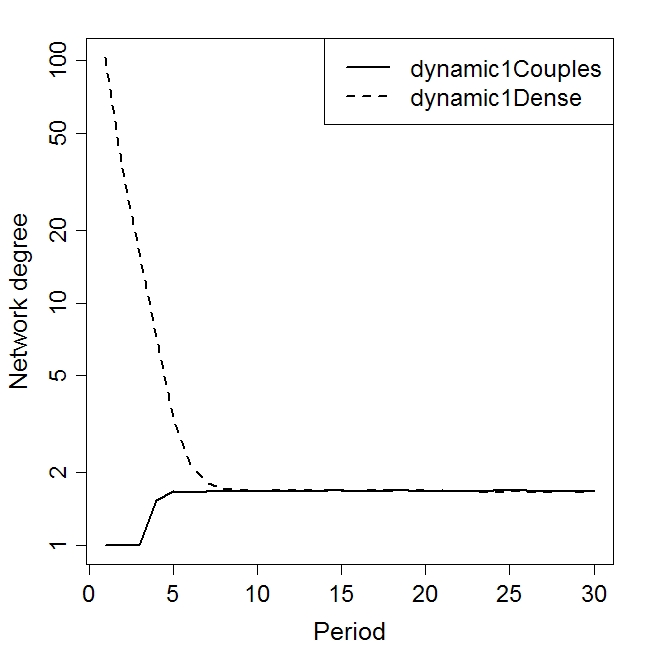}\includegraphics[width=.333\textwidth]{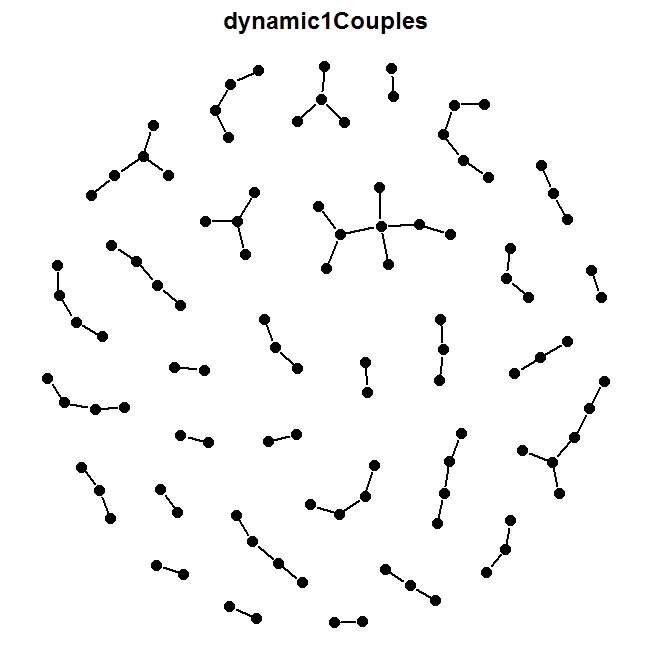}\includegraphics[width=.333\textwidth]{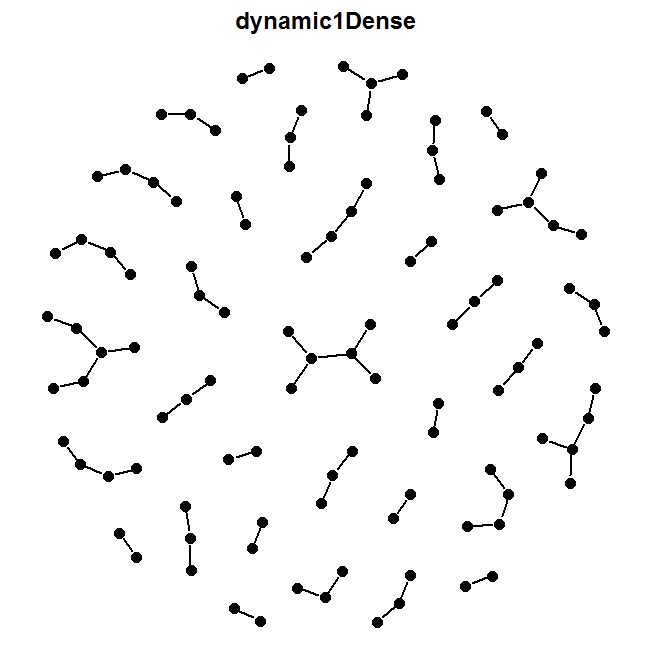}
\caption{The system resulting from the \emph{dynamic1} models converged to a fixed equilibrium independently from the starting point. The left panel shows the average degree of the networks resulting from the two models. The center and the right panel report the networks resulting after 30 periods of a typical run of the \emph{dynamic1Couples} and the \emph{dynamic1Dense} model respectively.}
\label{fig:dyn1}
\end{figure}

Especially interesting was the network dynamic. Apparently the system had a single equilibrium, which was reached independently from the starting point. From period 9 onwards, the average degree of the networks resulting from the \emph{dynamic1Couples} and the \emph{dynamic1Dense} models converged to the value of $1.67 \pm 0.02$ (Fig. \ref{fig:dyn1}, left panel). This was true although the initial degrees of the two networks were, respectively, 1 and 103. The final network structure of the two models was also very similar with, in both cases, all agents linked in small groups of 2-8 individuals (Fig. \ref{fig:dyn1}, center and right panels). 

Most of the times, the center of the group was represented by an especially cooperative agent, a fact consistent with the findings of \citet{Ehuiluz2005} model, where agents played a Prisoner's Dilemma on a dynamic network and the equilibrium was guaranteed  by ``leaders'' (agents in central positions) that played an essential role in sustaining cooperation in the system. In the network plotted in the middle of Figure \ref{fig:dyn1}, agents with more than two links had an average $\alpha_i$ of 4.02, while the one for agents with two links or less was 3.75. Similarly, agents with more than two links had an average $\beta_i$ of -- 0.21 (vs. 0.22) and an average $\gamma_i$ of 0.21 (vs. 0.13). More generally, both $\alpha_i$ and $\gamma_i$ of agents were positively correlated with the number of interactions in the last 10 periods of the game. In the \emph{dynamic1couples} the correlation coefficients were, respectively, 0.42 and 0.86, while in the \emph{dynamic1dense} they were 0.40 and 0.78.\footnote{All estimates were significant at the 1\% level.} In other words, agents with more links (thereby performing more interactions) were both more trustful and more trustworthy. 

As for the static models, we studied the relation between the agents' parameters and their earnings.  Table \ref{tab:payoffsDynamic} report  our estimations. The results of the \emph{dynamic1} models did not depart much from the static network ones. Again trustful and trustworthy agents were disadvantaged, while the capacity to react made little difference. One noticeable exception appeared in the final periods of the \emph{dynamic1Couples} model, where the coefficient related to the $\beta_i$ parameter became significant. This was rather surprising, because the small stable emerging groups (Fig. \ref{fig:dyn1}, center panel) were not very different from the dyads that characterized the \emph{fixedCouples} model, which also produced a significant correlation between $\beta_i$ and agents' earnings.

\begin{table}[!t]
	\centering
	\begin{tabular}{lrr@{}lr@{}lr@{}l}
	\toprule
	Model name & Period & \multicolumn{2}{c}{$\alpha_i$} & \multicolumn{2}{c}{$\beta_i$} & \multicolumn{2}{c}{$\gamma_i$} \\
	\midrule
 	\emph{dynamic1Couples} & 1--10 & --0.78 & $^{***}$ & 0.11 & & --080 & $^{***}$ \\
																			& 11--20 & --0.78 & $^{***}$ & 0.16 & & --0.67 & $^{***}$ \\
																			& 21--30 & --0.68 & $^{***}$ & 0.21 & $^{*}$ & --0.66 & $^{***}$ \\
	\midrule																								
 	\emph{dynamic1Dense} 	& 1--10 & 0.02 & & --0.18 & & 0.64 & $^{***}$ \\ 
 																			& 11--20 & --0.72 & $^{***}$ & 0.17 & & --0.66 & $^{***}$ \\
 																			& 21--30 & --0.66 & $^{***}$ & 0.15 & & --0.61 & $^{***}$ \\
	\midrule
 	\emph{dynamic2Couples} & 1--10  & --0.13 &  & --0.05 & & 0.53 & $^{***}$ \\
 																			& 11--20 & 0.37 & $^{***}$ & --0.14 & & 0.96 & $^{***}$ \\
 																			& 21--30 & 0.37 & $^{***}$ & --0.14 & & 0.95 & $^{***}$ \\	 	 	
	\midrule
 	\emph{dynamic2k10} & 1--10  & 0.32 & $^{***}$ & --0.19 & & 0.87 & $^{***}$ \\
 																& 11--20 & 0.36 & $^{***}$ & --0.18 & & 0.96 & $^{***}$ \\
 																& 21--30 & 0.35 & $^{***}$ & --0.19 & & 0.97 & $^{***}$ \\	 	
	\bottomrule			
	\end{tabular}
	\caption{Correlation between average payoffs per run and agents' parameters. Significance codes: $^{***} p < 0.001$, $^{**} p < 0.01$, $^{*} p < 0.05$.}
	\label{tab:payoffsDynamic}
\end{table}

Another exception was found in the initial periods of the \emph{dynamic1Dense} model, where the correlation coefficient for $\alpha_i$ was almost zero and the one for $\gamma_i$ was positive and highly significant. However, this situation did not persist and in the period 11--20 the coefficient became already positive and highly significant. The initial advantage for trustworthy agents was due to their capacity of maintaining their links longer, while untrustworthy lost theirs because of the unsatisfaction of their opponents. Similarly, trustful agents were not worse off than the others because they quickly broke their unsatisfying links. However, once  the network settled up, the situation changed and the coefficients became similar to the \emph{dynamic1Couples} model.

The \emph{dynamic2Couples} model led instead to a strong increase in cooperation (Tab.	\ref{tab:meansDynamic}). Returns were already significantly higher than in the experiment in the first 10 periods of the game and further increased subsequently. Investments were significantly higher than in the experiment from the 10$^{th}$ period onwards. Starting from the same period,  returns were higher than investments as well. In other words, investments became profitable for $A$ players. This was also clear when looking at the correlations presented in Table \ref{tab:payoffsDynamic}. The correlation between the agents' payoffs and both their $\alpha_i$ and $\gamma_i$ became \emph{positive} and significant, i.e., trustful and trustworthy agents earned, on average, higher payoffs. The explanation is that more trustful and trustworthy agents had more links, so enjoying a higher number of exchanges than less trustful/trustworthy ones. More precisely, the correlations between $\alpha_i$ and $\gamma_i$ of each agent and the number of interactions that the same agent performed during the whole 30 period game were both positive and significant ($r=0.38$, $p<0.001$ and $r=0.92$, $p<0.001$ respectively). Vice versa, less cooperative agents became easily isolated and earned, on average, lower payoffs.

\begin{figure}[!t]
\centering
\includegraphics[width=.333\textwidth]{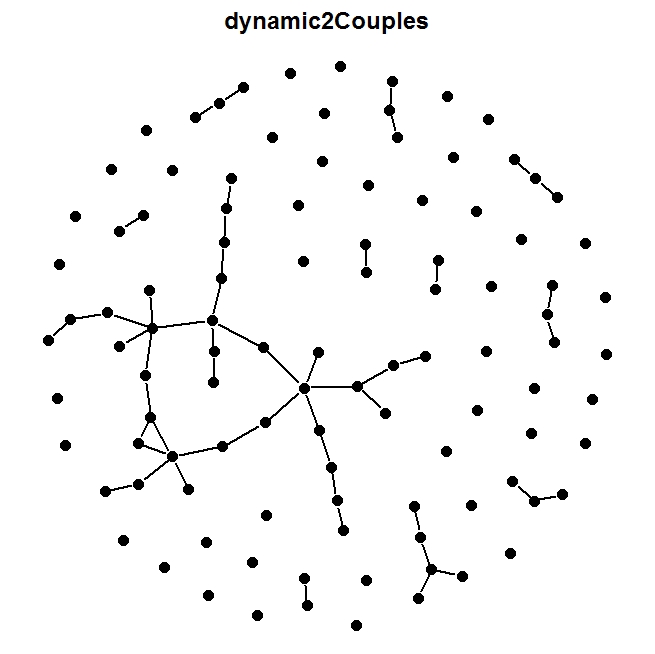}\hspace{3cm}\includegraphics[width=.333\textwidth]{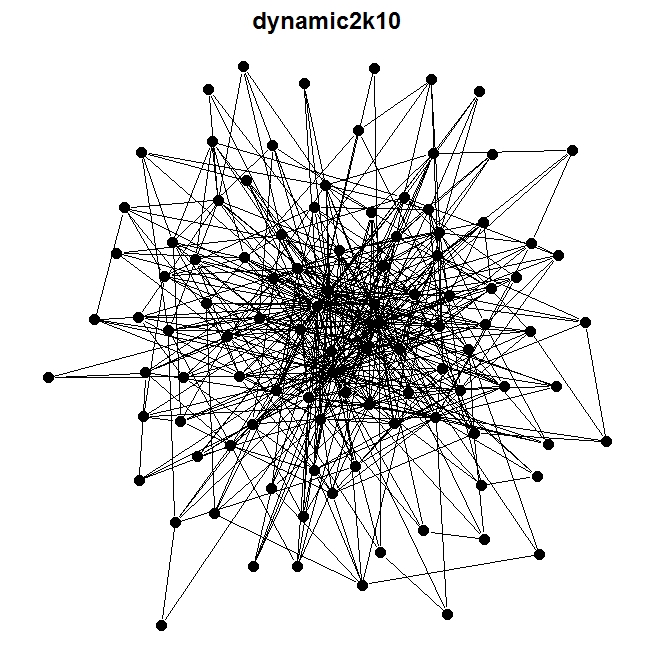}
\caption{Networks resulting after 30 periods of a typical run of the \emph{dynamic2Couples} model (left) and of the \emph{dynamic2k10} model (right).}
\label{fig:dyn2}
\end{figure}

The resulting network was also remarkably different from the \emph{dynamic1} model.  Many agents were now isolated, while a large cluster emerged around the most cooperative agents (Fig. \ref{fig:dyn2}). Therefore, the most trustful/trustworthy agents were the most connected. For instance, the average $\alpha_i$ of agents with zero or one link in Fig. \ref{fig:dyn2} was 3.41, while the average $\alpha_i$ of agents with two or more links was 4.57. Similarly, the average $\gamma_i$ of agents with zero or one link was 0.10, while the average $\gamma_i$ of agents with two of more links was $0.22$.

The \emph{dynamic2k10} model presented a strong increase of cooperation (Tab.	\ref{tab:meansDynamic}). Unlike before, both the average investments and returns were higher than in the experiment already in the first ten periods. Investments and returns further increased in the subsequent periods, even if they finally settled down at a slightly lower value than in the \emph{dynamic2Couples} model. As before, investments became profitable for $A$ player from period 10 onwards. The correlations between agents' parameters and payoffs followed a path similar to the \emph{dynamic2Couples} model, with the exception that the correlation between the agents' $\alpha_i$ and their payoffs was positive and significant from the beginning of the game (Tab. \ref{tab:payoffsDynamic}). Overall, trustful and trustworthy agents enjoyed a larger number of interactions. The correlation between agents' parameter and number of interactions was  significant at the 0.1\% level for both $\alpha_i$ and $\gamma_i$ (while it was negative and significant only at the 10\% level for $\beta_i$).

The network resulting from a typical run of the \emph{dynamic2k10} model was presented in Figure \ref{fig:dyn2} (right). Unlike the previous case, no agent was now isolated. However, cooperative agents maintained a larger number of links. For instance, the average $\alpha_i$ of agents with five or less links was 2.73, while for agents with more than five links this figure is 4.17. Similarly the average $\gamma_i$ of the former group was 0.07, while the one for the latter group is 0.17.

\section{Discussion} \label{sec:discussion}

In this paper, we have presented an experimentally grounded ABM that extended the results of lab experiments on trust to study the impact of the interaction structure on the cooperative outcomes. The first important lesson that we can draw from our results is that it is not the stable or unstable nature of the interaction structure \emph{per se} which has a relevant impact on trust diffusion and cooperation. Our result supported the hypothesis that what really matters is the link of interaction outcome and structure formation. While a simple change of the network structure did not alter significantly the outcome of our game, the introduction of a dynamic network led  to higher cooperation when cooperative agents had the capability to create more links and, possibly, to isolate free-riders.

This first result stands in stark contrast with the popular literature emphasizing the effects of network structure on interaction outcomes \citep[e.g.][]{Barabasi2009,Buchanan2002,Watts1999}. The ``shadow of the future'' postulated by \citet{Axelrod1984} and \citet{Cohen2001} certainly represents an important factor in explaining cooperative outcomes. However, once trust has been eroded in static networks by untrustworthy actions, the stability of links between agents using conditional strategies can only produce a never ending run of mutual defection. On the contrary, by permitting agents to leave their partners cooperation increases since  the exploitation opportunities of free-riders are reduced. This is exactly what happened in our simulations under the \emph{dynamic2} conditions. Trustful and trustworthy agents were no longer bound to reciprocate defection, but were in position to find new partners with whom they started cooperative relations. As a consequence, overall cooperation increased, bringing higher fairness and efficiency in the system.

This has implications for sociological and experimental analyses on cooperation and social norms. Previous simulation models  highlighted the relevance of tags for cooperation \citep[e.g.][]{Axelrod2004,Hales2000}. In sociological terms, tags can be viewed as communication signals that circulate among people and connotate certain attributes of individuals or groups.  For instance, they include speaking dialects or particular accents, dressing elegant formal get-ups, or driving garish customized sport-cars. Although largely unintentional and often inaccurate or even false, these attributes or actions might be socially used as predictors of the others' behavior, in particular when interacting with unknown partners, although they can be unintentional, inaccurate or false. As suggested by the simulation model presented in \citet{Riolo-etal2001}, the presence of tags in a social system can help cooperation to emerge even without past experience, memory, reciprocity motives or reputational signals. This makes partner selection different than in our case, where it was based on past experience of each agent individually. The point is that, if individuals can detect tags or exchange some communication signals before interacting, it is likely that the intelligibility of partner selection and the consequent structure formation can be strongly influenced. This is where the dark side of partner selection is likely to enter the picture, i.e., when tag-based selection is biased by discriminative social beliefs and gives rise to inefficient network formation by preventing unknown partners from potential interaction benefits. If this is reasonable, future laboratory experiments that compare the impact of social tags and past experience on partner selection and structure formation should be particularly welcomed.

Another point is that, if it is reasonable to expect that interaction outcomes matter to give rise to social structures, as we have stressed, it should be essential to investigate empirically and experimentally how social expectations about the trustworthiness of potential partners and the trusting-other attitudes of individuals are influenced by cultural aspects, institutions and normative values at a social level. This is of particular interest in the study on the nature and evolution of markets, as suggested by \citet{North2005}. Firstly, particular market institutional scaffolds might be more efficient that others because they have institutionalized partner selection as a relatively low-cost sanctioning for cheaters. Secondly, particular market structure configurations might have historically evolved out of partner selection mechanisms, as suggested also by \citet{White2004}, so that complex market structures could be nothing but a magnification of relatively simple social mechanisms..

A second lesson concerns precisely the relevance of partner selection.  The increase of cooperation via partner selection has implication both for the efficiency of the system and for its fairness. In the \emph{dynamic2} conditions the average investment increased by one third and the average return almost doubled in comparison with both the experiment and the static networks. While the standard in the experimental literature on trust and cooperation often assumed randomly re-matching couples, as though the game were a sum of one-shot periods, with no chance for partner selection, our results suggested that cooperation can crystallize in stable interaction structures that were likely to depend on the positive outcomes generated by agent interaction. This means that it is reasonable to suppose that individual action and interaction structure tend to co-evolve in the social life \citep[e.g.][]{Buskens2000,Buskens2008a,Mark1998}. Our experimentally grounded model allows us to emphasize that one of these self-reinforcing co-evolutionary links could be driven by partner selection \citep[e.g.][]{Corten2009,Flache2001,Ehuiluz2005}.  If this is reasonable, by excluding from the radar screen of the experimental analysis these aspects, it is likely that something crucial for explaining social mechanisms of cooperation gets lost.

The third lesson concerns the research methods. By combining laboratory experiments and simulation models, we were able to exploit some of the advantages of both. This is not a common practice, even if earlier works that followed our same inspiration have already suggested some of its potentialities \citep[e.g.][]{Boero2010,DalForno2004,Deadman2000,Jager2003,Rouchier2003}. Being empirically defined, the distribution of strategies used by our agents was more plausible than what was commonplace in research based on formal models. At the same time, thanks to simulation, we were able to extend the experimental results beyond the limits of the laboratory (e.g., a larger number of interactions, more complex network structures, etc.).  

The link between experiments and simulation is not limited to the creation of empirically-grounded models. Verifying in the lab hypotheses deriving from simulation research is becoming relatively  common. A recent example of experimentally tested ABM is \citet{Centola2005} ``Emperor's Dilemma'', whose conclusions have been tested by \citet{Willer2009}. Our simulation evidence can therefore guide the design of a new set of both field studies and laboratory  experiments explicitly targeted to study partner selection behavior or, more generally, to gain a better understanding of the relation between individual preferences and the dynamics of social networks.   For instance,  our model supports the hypothesis that allowing individuals to select their interaction partners increases \emph{per se}  the amount of both trust and trustworthiness in a given system. This could be checked in field studies by comparing contexts of social interaction that vary in their degree of individual freedom in choosing their partners (e.g., firm and consumer behavior in monopolistic vs. competitive markets). It is also reasonable to suppose that partner selection can cover relevant functions of informal social control and sanctioning in many social and economic situations when the partner choice and the break of existing links becomes a socially shared information. This opens the door to interesting investigations on the relations between partner selection and reputation both for social order, as a means for norm abiders to isolate and punish cheaters \citep[e.g.][]{Pinyol2008}, and for environmental exploration, as a means to know what is the best partner choice in uncertain situations \citep[e.g.][]{Boero2010}.
The same hypothesis could be checked with more accuracy in the lab by designing a trust-game experiment where individuals can decide whether maintaining or not their current relations, even if this comes at the cost of reducing the number of interactions and the range of potential partners.

To sum up, by linking trust and partner selection our results emphasize new research directions. What is especially important is that the level of analysis moves from static to dynamic networks. Real world networks are rarely fixed and independent of individual actions. They  evolve  over time as the result of a myriad of independent choices done by agents embedded therein. Taking into account the aggregate level effects of these choices is complex and will plausibly involve a great deal of both empirical and computational new research. However, this will nicely improve our knowledge of fundamental social mechanisms and, \emph{a fortiori}, of important social situations.

 \paragraph{Aknowledgements} 
Financial support was provided by a FIRB 2003 grant (Strategic Program on Human, Economic and Social Sciences) from the Italian Ministry for the University and the Scientific Research (SOCRATE Research Project, Protocol: RBNE03Y338\_ 002) and by the Regione Piemonte, Research Project ICT4Law. A preliminary version of this paper has been presented at the Sixth European Social Simulation Conference, held in 2009 in Guilford, Surrey, UK. We thank three anonymous conference reviewers and the audience for useful remarks. The authors gratefully acknowledge comments and suggestions by Jakob Grazzini, Cristina Mosso  and K\'{a}roly Tak\'{a}cs.

\bibliographystyle{authordate1}
\bibliography{c:/Documenti/Bibliografie/GB}

\begin{thebibliography}{}

\bibitem[\protect\citename{Aktipis, }2004]{Aktipis2004}
Aktipis, A. 2004.
\newblock Know When to Walk Away: Contingent Movement and the Evolution of
  Cooperation.
\newblock {\em Journal of Theoretical Biology}, {\bf 231}, 249--260.

\bibitem[\protect\citename{Ashlock {\em et~al.\ }\relax, }1996]{Ashlock1996}
Ashlock, D., Smucker, M.~D., Stanley, E.~A., \& Tesfatsion, L. 1996.
\newblock Preferential Partner Selection in an Evolutionary Study of Prisoner's
  Dilemma.
\newblock {\em BioSystems}, {\bf 37}, 99--125.

\bibitem[\protect\citename{Axelrod {\em et~al.\ }\relax, }2002]{Axelrod2002}
Axelrod, R., Riolo, R., \& Cohen, M.~D. 2002.
\newblock Beyond Geography: Cooperation with Persistent Links in the Absence of
  Clustered Neighborhood.
\newblock {\em Personality and Social Psychology Review}, {\bf 6}(4), 341--346.

\bibitem[\protect\citename{Axelrod {\em et~al.\ }\relax, }2004]{Axelrod2004}
Axelrod, R., Hammond, R.~A., \& Grafen, A. 2004.
\newblock Altruism via Kin Selection: Strategies that Rely on Arbitrary Tags
  with which They Evolve.
\newblock {\em Evolution}, {\bf 58}(8), 1833--1838.

\bibitem[\protect\citename{Axelrod, }1984]{Axelrod1984}
Axelrod, Robert. 1984.
\newblock {\em The Evolution of Cooperation}.
\newblock New York: Basic Books.

\bibitem[\protect\citename{Axelrod, }1997]{Axelrod1997}
Axelrod, Robert. 1997.
\newblock {\em The Complexity of Cooperation: Agent-Based Models of Competition
  and Collaboration}.
\newblock Princeton: Princeton University Press.

\bibitem[\protect\citename{Back, }2010]{Back2010}
Back, I.~H. 2010.
\newblock Commitment Bias: Mistaken Partner Selection or Ancient Wisdom?
\newblock {\em Evolution and Human Behavior}, {\bf 31}, 22--28.

\bibitem[\protect\citename{Back \& Flache, }2006]{Back2006}
Back, I.~H., \& Flache, A. 2006.
\newblock The Viability of Cooperation Based on Interpersonal Commitment.
\newblock {\em Journal of Artificial Societies and Social Simulation}, {\bf
  9}(1), 12.

\bibitem[\protect\citename{Back \& Flache, }2008]{Back2008}
Back, I.~H., \& Flache, A. 2008.
\newblock The Adaptive Rationality of Interpersonal Commitment.
\newblock {\em Rationality and Society}, {\bf 20}(1), 65--83.

\bibitem[\protect\citename{Barab\'{a}si \& Albert, }1999]{Barabasi1999}
Barab\'{a}si, Albert-L\'{a}szl\'{o}, \& Albert, Réka. 1999.
\newblock Emergence of Scaling in Random Networks.
\newblock {\em Science}, {\bf 289}, 509--512.

\bibitem[\protect\citename{Barabási, }2009]{Barabasi2009}
Barabási, Albert-László. 2009.
\newblock Scale-Free Networks: A Decade and Beyond.
\newblock {\em Science}, {\bf 325}, 412--413.

\bibitem[\protect\citename{Barrera \& Buskens, }2009]{Barrera2009}
Barrera, Davide, \& Buskens, Vincent. 2009.
\newblock Third-Party Effects on Trust in an Embedded Investment Game.
\newblock {\em In:} Cook, K., Snijders, C., Buskens, V., \& Cheshire, C. (eds),
  {\em Trust and Reputation}.
\newblock Russell Sage: New York.
\newblock Forthcoming.

\bibitem[\protect\citename{Beckman {\em et~al.\ }\relax, }2004]{Beckman2004}
Beckman, C.~M., Haunschild, P.~R., \& Phillips, D.~J. 2004.
\newblock Friends or Strangers? Firm-Specific Uncertainty, Market Uncertainty,
  and Network Partner Selection.
\newblock {\em Organization Science}, {\bf 15}(3), 259--275.

\bibitem[\protect\citename{Berg {\em et~al.\ }\relax, }1995]{BDM1995}
Berg, Joyce, Dickhaut, John, \& McCabe, Kevin~A. 1995.
\newblock Trust, Reciprocity and Social History.
\newblock {\em Games and Economic Behavior}, {\bf 10}, 122--142.

\bibitem[\protect\citename{Boero \& Squazzoni, }2005]{BS2005}
Boero, Riccardo, \& Squazzoni, Flaminio. 2005.
\newblock Does Empirical Embeddedness Matter? Methodological Issues on
  Agent-Based Models for Analytical Social Science.
\newblock {\em Journal of Artificial Societies and Social Simulation}, {\bf
  8}(4), 6.

\bibitem[\protect\citename{Boero {\em et~al.\ }\relax, }2009a]{Boero2009c}
Boero, Riccardo, Bravo, Giangiacomo, Castellani, Marco, Laganà, Francesco, \&
  Squazzoni, Flaminio. 2009a.
\newblock Pillars of Trust: An Experimental Study on Reputation and Its
  Effects.
\newblock {\em Sociological Research Online}, {\bf 14}(5), 5.

\bibitem[\protect\citename{Boero {\em et~al.\ }\relax, }2009b]{Boero2009}
Boero, Riccardo, Bravo, Giangiacomo, Castellani, Marco, \& Squazzoni, Flaminio.
  2009b.
\newblock Reputational cues in repeated trust games.
\newblock {\em Journal of Socio-Economics}, {\bf 38}(6), 871--877.

\bibitem[\protect\citename{Boero {\em et~al.\ }\relax, }2010]{Boero2010}
Boero, Riccardo, Bravo, Giangiacomo, Castellani, Marco, \& Squazzoni, Flaminio.
  2010.
\newblock Why Bother with What Others Tell You? An Experimental Data-Driven
  Agent-Based Model.
\newblock {\em Journal of Artificial Societies and Social Simulation}, {\bf
  13}(2), 6.

\bibitem[\protect\citename{Bravo, }2010]{Bravo2010}
Bravo, Giangiacomo. 2010.
\newblock Agents' beliefs and the evolution of institutions for common-pool
  resource management.
\newblock {\em Rationality and Society}, {\bf 22}(3), In press.

\bibitem[\protect\citename{Buchanan, }2002]{Buchanan2002}
Buchanan, Mark. 2002.
\newblock {\em Nexus: small worlds and the groundbreaking science of networks}.
\newblock New York: W.W. Norton.

\bibitem[\protect\citename{Buskens \& Weesie, }2000]{Buskens2000}
Buskens, V., \& Weesie, J. 2000.
\newblock Cooperation via Social Networks.
\newblock {\em Analyse and Kritik}, {\bf 22}, 44--74.

\bibitem[\protect\citename{Buskens {\em et~al.\ }\relax, }2008]{Buskens2008a}
Buskens, V., Corten, R., \& Weesie, J. 2008.
\newblock Consent or Conflict: Coevolution of Coordination and Networks.
\newblock {\em Journal of Peace Research}, {\bf 45}, 205--222.

\bibitem[\protect\citename{Buskens \& Raub, }2008]{Buskens2008}
Buskens, Vincent, \& Raub, Werner. 2008.
\newblock Rational choice research on social dilemmas: embeddedness effects on
  trust.
\newblock {\em In:} Wittek, R., Snijders, T.A.B., \& Nee, V. (eds), {\em
  Handbook of Rational Choice Social Research}.
\newblock New York: Russell Sage.

\bibitem[\protect\citename{Calv\'{o}-Armengol, }2001]{Calvo2001}
Calv\'{o}-Armengol, A. 2001.
\newblock On Bargaining Partner Selection When Communication Is Restricted.
\newblock {\em International Journal of Game Theory}, {\bf 30}, 503--515.

\bibitem[\protect\citename{Camerer, }2003]{Camerer2003}
Camerer, Colin~F. 2003.
\newblock {\em Behavioral Game Theory. Experiments in Strategic Interaction}.
\newblock New York / Princeton: Russel Sage Foundation / Princeton University
  Press.

\bibitem[\protect\citename{Centola {\em et~al.\ }\relax, }2005]{Centola2005}
Centola, Damon, Willer, Robb, \& Macy, Michael. 2005.
\newblock The Emperor’s Dilemma: A Computational Model of Self-Enforcing Norms.
\newblock {\em American Journal of Sociology}, {\bf 110}(4), 1009--40.

\bibitem[\protect\citename{Chmura \& Pitz, }2007]{Chmura2007}
Chmura, T., \& Pitz, T. 2007.
\newblock An Extended Reinforcement Algorithm for Estimation of Human Behaviour
  in Experimental Congestion Games.
\newblock {\em Journal of Artificial Societies and Social Simulation}, {\bf
  10}(2), 1.

\bibitem[\protect\citename{Cohen {\em et~al.\ }\relax, }2001]{Cohen2001}
Cohen, M., Riolo, R., \& Axelrod, R. 2001.
\newblock The Role of Social Structure in the Maintenance of Cooperative
  Regimes.
\newblock {\em Rationality and Society}, {\bf 13}, 5--32.

\bibitem[\protect\citename{Coleman, }1990]{Coleman1990}
Coleman, James~S. 1990.
\newblock {\em Foundations of Social Theory}.
\newblock Harvard: Harvard University Press.

\bibitem[\protect\citename{Corten \& Buskens, }2009]{Corten2009}
Corten, Rense, \& Buskens, Vincent. 2009.
\newblock {\em Co-evolution of Conventions and Networks: An Experimental
  Study}.
\newblock ISCORE Paper 44:
  <http://www.uu.nl/uupublish/content/CortenBuskens\_ISCORE.pdf>.

\bibitem[\protect\citename{Cronk, }2007]{Cronk2007}
Cronk, Lee. 2007.
\newblock The influence of cultural framing on play in the trust game: a Maasai
  example.
\newblock {\em Evolution and Human Behavior}, {\bf 28}, 352--358.

\bibitem[\protect\citename{Dal~Forno \& Merlone, }2004]{DalForno2004}
Dal~Forno, A., \& Merlone, U. 2004.
\newblock From Classroom Experiments to Computer Code.
\newblock {\em Journal of Artificial Societies and Social Simulation}, {\bf
  7}(2), 2.

\bibitem[\protect\citename{De~Vos {\em et~al.\ }\relax, }2001]{DeVos2001}
De~Vos, H., Smaniotto, R., \& Elsas, D.~A. 2001.
\newblock Reciprocal Altruism under Conditions of Partner Selection.
\newblock {\em Rationality and Society}, {\bf 13}, 5--32.

\bibitem[\protect\citename{Deadman {\em et~al.\ }\relax, }2000]{Deadman2000}
Deadman, Peter, Schlager, Edella, \& Gimblett, Randy. 2000.
\newblock Simulating Common Pool Resource Management Experiments with Adaptive
  Agents Employing Alternate Communication Routines.
\newblock {\em Journal of Artificial Societies and Social Simulation}, {\bf
  3}(2), 2.

\bibitem[\protect\citename{Dutta {\em et~al.\ }\relax, }2005]{Dutta2005}
Dutta, B., Ghosal, S., \& D., Ray. 2005.
\newblock Farsighted Network Formation.
\newblock {\em Journal of Economic Theory}, {\bf 122}, 143--164.

\bibitem[\protect\citename{Ebenh\"oh \& Pahl-Wostl, }2008]{Ebenhoh2008}
Ebenh\"oh, Eva, \& Pahl-Wostl, Claudia. 2008.
\newblock Agent Behavior Between Maximization and Cooperation.
\newblock {\em Rationality and Society}, {\bf 20}(3), 227--252.

\bibitem[\protect\citename{Egu\'{i}luz {\em et~al.\ }\relax,
  }2005]{Ehuiluz2005}
Egu\'{i}luz, V\'{i}ctor~M., Zimmermann, Mart\'{i}n~G., Cela-Conde, Camilo J.
  Cela-Conde, \& San~Miguel, Maxi San~Miguel. 2005.
\newblock Cooperation and the Emergence of Role Differentiation in the Dynamics
  of Social Networks.
\newblock {\em American Journal of Sociology}, {\bf 110}(4), 977--1008.

\bibitem[\protect\citename{Fischbacher, }2007]{Fischbacher2007}
Fischbacher, Urs. 2007.
\newblock z-Tree: Zurich toolbox for ready-made economic experiments.
\newblock {\em Experimental Economics}, {\bf 10}, 171--178.

\bibitem[\protect\citename{Flache, }2001]{Flache2001}
Flache, Andreas. 2001.
\newblock Individual Risk Preferences and Collective Outcomes in the Evolution
  of Exchange Networks.
\newblock {\em Rationality and Society}, {\bf 13}, 304--348.

\bibitem[\protect\citename{Gulati, }1995]{Gulati1995}
Gulati, R. 1995.
\newblock Performance, Aspirations and Risky Organizational Change.
\newblock {\em Administrative Science Quarterly}, {\bf 43}, 58--86.

\bibitem[\protect\citename{Hales, }2000]{Hales2000}
Hales, D. 2000.
\newblock Cooperation without Space or Memory: Tags, Groups and the Prisoner's
  Dilemma.
\newblock {\em Pages  157--166 of:} Moss, S., \& Davidsson, P. (eds), {\em
  Multi-Agent Based Simulation}.
\newblock Berlin-Heidelberg: Springer Verlag.

\bibitem[\protect\citename{Haruvy {\em et~al.\ }\relax, }2006]{Haruvy2006}
Haruvy, E., Roth, A., \& \"{U}nver, M. 2006.
\newblock The Dynamics of Law-Clerk Matching: An Experimental and Computational
  Investigation of Proposals for Reform of the Market.
\newblock {\em Journal of Economic Dynamics and Control}, {\bf 30}(3),
  457--486.

\bibitem[\protect\citename{Hauk, }2001]{Hauk2001}
Hauk, Esther. 2001.
\newblock Leaving the Prison: Permitting Partner Choice and Refusal in
  Prisoner's \ Dilemma Games.
\newblock {\em Computational Economics}, {\bf 18}, 65--87.

\bibitem[\protect\citename{Jackson \& Watts, }2002]{Jackson2002}
Jackson, M.~O., \& Watts, A. 2002.
\newblock The Evolution of Social and Economic Networks.
\newblock {\em Journal of Economic Theory}, {\bf 106}, 265--295.

\bibitem[\protect\citename{Jackson \& Wolinksy, }1996]{Jackson1996}
Jackson, M.~O., \& Wolinksy, A. 1996.
\newblock A Strategic Model of Social and Economic Networks.
\newblock {\em Journal of Economic Theory}, {\bf 71}, 44--74.

\bibitem[\protect\citename{Jager \& Janssen, }2003]{Jager2003}
Jager, Wander, \& Janssen, Marco~A. 2003.
\newblock Using Artificial Agents to Understand Laboratory Experiments of
  Common-Pool Resources with Real Agents.
\newblock {\em Pages  75--102 of:} Janssen, Marco~A. (ed), {\em Complexity and
  Ecosystem Management: The Theory and Practice of Multi-Agent Systems}.
\newblock Cheltenham: Edward Elgar.

\bibitem[\protect\citename{Janssen \& Ostrom, }2006]{Janssen2006d}
Janssen, Marco~A., \& Ostrom, Elinor. 2006.
\newblock Empirically Based, Agent-based models.
\newblock {\em Ecology and Society}, {\bf 11}(2), 37.

\bibitem[\protect\citename{Joyce {\em et~al.\ }\relax, }2006]{Joyce-etal2006}
Joyce, David, Kennison, John, Densmore, Owen, Guerin, Steven, Barr, Shawn,
  Charles, Eric, \& Thompson, Nicholas~S. 2006.
\newblock My Way or the Highway: a More Naturalistic Model of Altruism Tested
  in an Iterative Prisoners' Dilemma.
\newblock {\em Journal of Artificial Societies and Social Simulation}, {\bf 9},
  4.

\bibitem[\protect\citename{Kagel \& Roth, }2000]{Kagel2000}
Kagel, J., \& Roth, A. 2000.
\newblock The Dynamics of Reorganization in Matching Markets: A Laboratory
  Experiment Motivated by a Natural Experiment.
\newblock {\em Quarterly Journal of Economics}, {\bf 115}, 201--235.

\bibitem[\protect\citename{Kahneman \& Tversky, }2000]{KT2000}
Kahneman, Daniel, \& Tversky, Amos (eds). 2000.
\newblock {\em Choices, Values and Frames}.
\newblock Cambridge: Cambridge University Press / Russell Sage Foundation.

\bibitem[\protect\citename{Keser, }2003]{Keser2003}
Keser, Claudia. 2003.
\newblock Experimental Games for the Design of Reputation Management Systems.
\newblock {\em IBM Systems Journal}, {\bf 42}, 498--506.

\bibitem[\protect\citename{King-Casas {\em et~al.\ }\relax,
  }2005]{King-Casas-etal2005}
King-Casas, Brooks, Tomlin, Damon, Anen, Cedric, Camerer, Colin~F., \&
  Montague, Steven R. Quartzand P.~Read. 2005.
\newblock Getting to Know You: Reputation and Trust in a Two-Person Economic
  Exchange.
\newblock {\em Science}, {\bf 308}, 79--83.

\bibitem[\protect\citename{Knoch {\em et~al.\ }\relax, }2009]{Knoch2009}
Knoch, Daria, Schneider, Frederic, Schunk, Daniel, Hohmann, Martin, \& Fehr,
  Ernst. 2009.
\newblock Disrupting the prefrontal cortex diminishes the human ability to
  build a good reputation.
\newblock {\em Proceedings of the National Academy of Sciences USA}, {\bf
  106}(49), 20895--20899.

\bibitem[\protect\citename{Kollock, }1994]{Kollock1994}
Kollock, Peter. 1994.
\newblock The Emergence of Exchange Structures: An Experimental Study of
  Uncertainty, Commitment, and Trust.
\newblock {\em American Journal of Sociology}, {\bf 100}(2), 313--345.

\bibitem[\protect\citename{Mark, }1998]{Mark1998}
Mark, N. 1998.
\newblock Beyond Individual Differences: Social Differentiation from First
  Principles.
\newblock {\em American Sociological Review}, {\bf 63}, 309--330.

\bibitem[\protect\citename{Molm {\em et~al.\ }\relax, }2000]{Molm2000}
Molm, L.~D., Takahashi, N., \& Peterson, G. 2000.
\newblock Risk and Trust in Social Exchange: An Experimental Test of a
  Classical Proposition.
\newblock {\em American Journal of Sociology}, {\bf 105}(5), 1396--1427.

\bibitem[\protect\citename{North, }2005]{North2005}
North, Douglass~C. 2005.
\newblock {\em Understanding the Process of Economic Change}.
\newblock Princeton, NJ: Princeton University Press.

\bibitem[\protect\citename{Ortmann {\em et~al.\ }\relax, }2000]{Ortmann2000}
Ortmann, A., Fitzgerald, J., \& Boeing, C. 2000.
\newblock Trust, Reciprocity, and Social History: {A} Re-examination.
\newblock {\em Experimental Economics}, {\bf 3}, 81--100.

\bibitem[\protect\citename{Pinyol {\em et~al.\ }\relax, }2008]{Pinyol2008}
Pinyol, I., Paolucci, F., Sabater-Mir, J., \& R., Conte. 2008.
\newblock Beyond Accuracy. Reputation for Partner Selection with Lies and
  Retaliation.
\newblock {\em Pages  128--140 of:} Antunes, L., Paolucci, M., \& Norling, E.
  (eds), {\em Multi-Agent-Based Simulation {VIII}}.
\newblock Berlin-Heidelberg: Springer Verlag.

\bibitem[\protect\citename{Podolny, }2001]{Podolny2001}
Podolny, J. 2001.
\newblock Networks as the Pipes and Prisms of the Market.
\newblock {\em American Journal of Sociology}, {\bf 107}, 33--60.

\bibitem[\protect\citename{Pujol {\em et~al.\ }\relax, }2005]{Pujol2005}
Pujol, J.~M., Flache, A., Delgado, J., \& Sanguesa, R. 2005.
\newblock How Can Social Networks Ever Become Complex? {Modelling} the
  Emergence of Complex Networks from Local Social Exchanges.
\newblock {\em Journal of Artificial Societies and Social Simulation}, {\bf
  8}(4), 12.

\bibitem[\protect\citename{{R Development Core Team}, }2009]{R2009}
{R Development Core Team}. 2009.
\newblock {\em R: A Language and Environment for Statistical Computing}.
\newblock R Foundation for Statistical Computing, Vienna, Austria.
\newblock {ISBN} 3-900051-07-0.

\bibitem[\protect\citename{Rauhut \& Junker, }2009]{Rauhut2009}
Rauhut, H., \& Junker, M. 2009.
\newblock Punishment Deters Crime Because Humans are Bounded in Their Strategic
  Decision-Making.
\newblock {\em Journal of Artificial Societies and Social Simulation}, {\bf
  12}(3), 1.

\bibitem[\protect\citename{Riolo {\em et~al.\ }\relax, }2001]{Riolo-etal2001}
Riolo, Rick~L., Cohen, Michael~D., \& Axelrod, Robert. 2001.
\newblock Evolution of Cooperation Without Reciprocity.
\newblock {\em Nature}, {\bf 414}, 441--443.

\bibitem[\protect\citename{Rouchier, }2003]{Rouchier2003}
Rouchier, J. 2003.
\newblock Re-Implementation of a Multi-Agent Model Aimed at Sustaining
  Experimental Economic Research: The Case of Simulations with Emerging
  Speculation.
\newblock {\em Journal of Artificial Societies and Social Simulation}, {\bf
  6}(4), 7.

\bibitem[\protect\citename{Slonim \& Garbarino, }2008]{Slonim2008}
Slonim, R., \& Garbarino, E. 2008.
\newblock Increases in Trust and Altruism from Partner Selection: Experimental
  Evidence.
\newblock {\em Experimental Economics}, {\bf 11}, 134--153.

\bibitem[\protect\citename{Watts, }1999]{Watts1999}
Watts, Duncan~J. 1999.
\newblock Networks, Dynamics and the Small-World Phenomenon.
\newblock {\em American Journal of Sociology}, {\bf 105}(2), 493--526.

\bibitem[\protect\citename{Watts \& Strogatz, }1998]{Watts1998}
Watts, Duncan~J., \& Strogatz, Steven~H. 1998.
\newblock Collective dynamics of ‘small-world’ networks.
\newblock {\em Nature}, {\bf 393}, 440--442.

\bibitem[\protect\citename{White, }2004]{White2004}
White, Harrison~C. 2004.
\newblock {\em Markets from Networks: Socioeconomic Models of Production}.
\newblock Princeton: Princeton University Press.

\bibitem[\protect\citename{Willer {\em et~al.\ }\relax, }2009]{Willer2009}
Willer, Robb, Kuwabara, Ko, \& Macy, Michael~W. 2009.
\newblock The False Enforcement of Unpopular Norms.
\newblock {\em American Journal of Sociology}, {\bf 115}(2), 451--490.

\bibitem[\protect\citename{Yagamashi {\em et~al.\ }\relax,
  }1998]{Yagamashi1998}
Yagamashi, T., Cook, K.~S., \& Watabe, M. 1998.
\newblock Uncertainty, Trust, and Commitment Formation in the {United States}
  and {Japan}.
\newblock {\em American Journal of Sociology}, {\bf 104}(1), 165--194.

\end{thebibliography}

\end{document}